\documentclass[12pt,twoside]{article}
\usepackage{fleqn,espcrc1}

\usepackage{graphicx,epsf}

\newcommand{\AmS}{{\protect\the\textfont2
  A\kern-.1667em\lower.5ex\hbox{M}\kern-.125emS}}

\title{The Continuum Spectrum of the 4N System. Results and Challenges}

\author{J. Carbonell\address{Institut des Sciences Nucl\'eaires, \\ 
        53 Av. des Martyrs, 38026 Grenoble, France}}%

\begin{document}

\maketitle

%

\section{INTRODUCTION}

The theoretical description of the A=4 scattering states constitutes
a serious challenge for the existing NN interaction models.
The reason for that is not purely technical, but lies rather
in the richness of the continuum spectrum itself.
Though far from the heavy nuclei imbroglio, it is the simplest system
which presents the main characteristics -- thresholds and resonances -- of the nuclear complexity.
Furthermore, they appear already in the low energy region 
which can be unlikely affected by the three nucleon forces, a keystone in the success
encountered when describing the A=3 states and A=4 bound state \cite{GK_NPA_93,VKR_FBS_95,GWHKG_96,PPCPW_97}.
 
Solutions of the 4N scattering states have been recently obtained
by different groups \cite{VRK_PRL_98,CCG_PLB_99,F_PRL_99}, solving Schr\"odinger, 
Faddeev-Yakubovsky or AGS equations, with realistic NN potentials. 

The aim of this contribution is
to give an overview of the main theoretical results existing for the A=4 scattering,
specially those obtained since the last Groningen Few-Body Conference.
A more detailed review including references can be found in \cite{CS_98}.

\section{n-$^3$H}

\begin{figure}[htb]
\vspace{-0.50cm}
\begin{minipage}[t]{77mm}
\epsfxsize=7.7cm\epsfysize=7.5cm\mbox{\epsffile{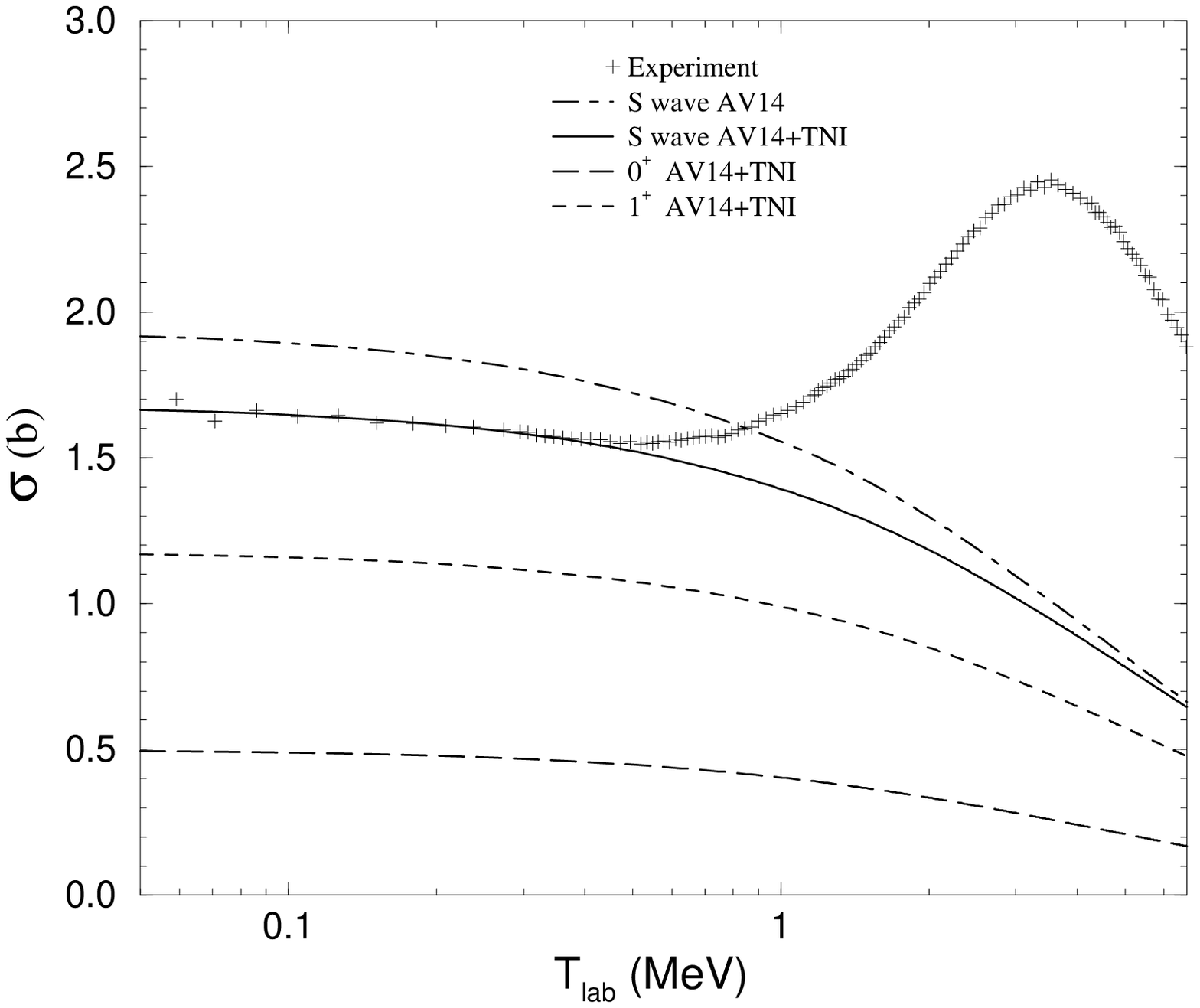}}
\vspace{-1.2cm}
\caption{Experimental n-$^3$H cross section and calculated S-wave contribution with (solid)
and without (dotted-dashed) TNI}\label{nt_S}
\end{minipage}
\hspace{0.5cm}
\begin{minipage}[t]{77mm}
\epsfxsize=7.7cm\epsfysize=7.5cm\mbox{\epsffile{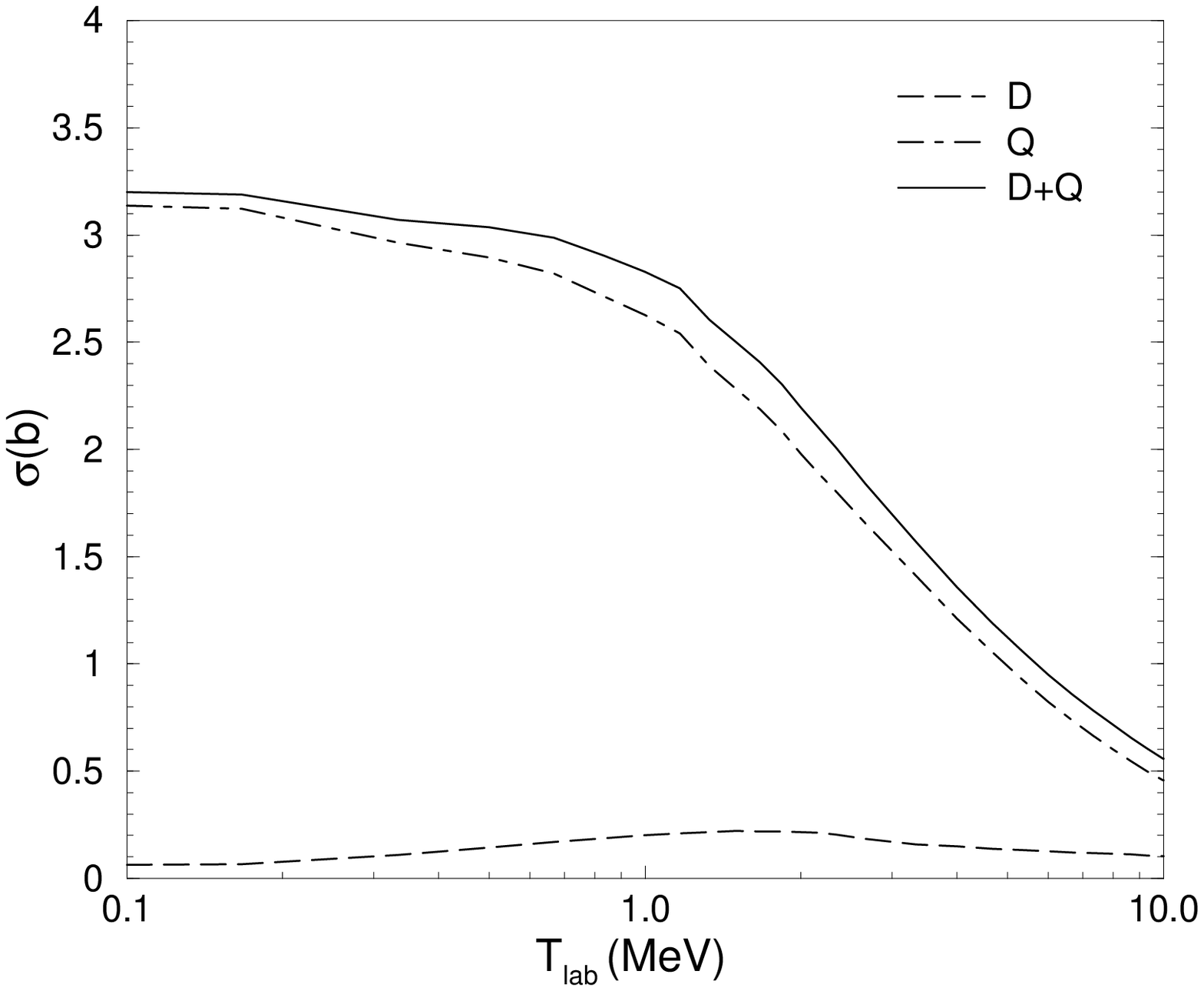}}
\vspace{-1.2cm}
\caption{Doublet and Quartet contribution to n-d elastic cross section}\label{nd}
\end{minipage}
\vspace{-0.50cm}
\end{figure}

The first topic concerns n+$^3$H,
the simplest A=4 system after the $\alpha$ particle. 
It is a pure  T=1 isospin state free from the Coulomb problems.
A simple look into its cross section and its
comparison with the n-d  case (Figures \ref{nt_S} and \ref{nd}) illustrate
well  the qualitative difference with respect the A=3 case.


The n-$^3$H scattering lengths with realistic potentials 
(AV14 \cite{AV14_PRC_84}, AV18 \cite{AV18_PRC_95}, Nijm II \cite{NIJ_PRC_93}) were presented 
in the last Few-Body Groningen Conference \cite{CCG_NPA_98,V_NPA_98}.
The singlet $a_0$ and triplet $a_1$ values (in fm) are summarized in the upper half part of Table~\ref{A_nt} together
with the deduced coherent scattering length $a_c={1\over4}a_0+{3\over4}a_1$  and the zero energy 
cross section $\sigma(0)=\pi(a_0^2+3a_1^2)$ in fm$^2$.

\begin{table}[htb]
\caption{n-t scattering length.}\label{A_nt}
\newcommand{\m}{\hphantom{$-$}}
\newcommand{\cc}[1]{\multicolumn{1}{c}{#1}}
\renewcommand{\tabcolsep}{0.8pc} 
\renewcommand{\arraystretch}{1.0} 
\begin{tabular}{@{}ll ll cc l}\hline
NN      &NNN          & $a_0$ & $a_1$ & $a_c$    & $\sigma(0)$  & Ref.   				  \\\hline
AV14    & ---         & 4.31  & 3.79  &  3.92	 & 194 	    & \cite{CCG_NPA_98,CCG_PLB_99} \\
        &             & 4.32  & 3.80  &  3.93	 & 195 	    & \cite{V_NPA_98,VRK_PRL_98}	  \\
AV18    & ---         & 4.32  & 3.76  &  3.90	 & 192 	    & \cite{V_NPA_98,VRK_PRL_98}	  \\
Nijm II & ---         & 4.31  & 3.76  &  3.90	 & 192 	    & \cite{CCG_NPA_98,CCG_PLB_99} \cr\hline
%
AV14    & Hyperadial  & 4.00  & 3.53  & 3.65     & 168          & \cite{CCG_NPA_98,CCG_PLB_99} \\ 
AV14    & Urbana VIII & 4.08  & 3.59  & 3.71     & 174          & \cite{V_NPA_98,VRK_PRL_98}\\ 
AV18    & Urbana IX   & 4.05  & 3.58  & 3.71     & 172          & \cite{V_NPA_98,VRK_PRL_98}\\ 
MT I-III& ---         & 4.10  & 3.63  & 3.75     & 177 	    & \cite{CC_PRC_98}    \\ \hline
Exp    &              &       &       &          & 170$\pm$3    & \cite{PBS_PRC_80}    \\ 
\end{tabular}
\vspace{-.5cm}
\end{table} 

One can see  on one hand similar ($<1\%$) values for different realistic potentials  
and on another hand a very good agreement using different methods.
In \cite{CCG_PLB_99,CCG_NPA_98,CC_PRC_98} the Faddeev-Yakubovsky (FY) equations in configuration space were solved whereas
authors of \cite{V_NPA_98,VRK_PRL_98} used the Correlated Hyperspherical Method (CHH).  
The comparison with the experimental cross section -- $\sigma(0)=170\pm3$ mb from \cite{PBS_PRC_80} --
shows that NN realistic potentials fail in describing the zero energy cross section as 
they fail in reproducing the three- ($B_3$) and four-nucleon ($B_4$) binding energies.  
Unlike the n-d case, the 4N scattering states call for three nucleon interaction (TNI) from the very beginning.
Indeed in the n-$^3$H case the singlet and triplet contribution are
of the same size (Figure~\ref{nt_S}) 
whereas the doublet n-d value -  directly correlated to $B_3$ and so affected by TNI -
turns to be one order of magnitude smaller than the quartet and has no visible effect in the
total cross section (Figure~\ref{nd}).
This smartness of nature made the inclusion of TNI unnecessary to reproduce the low energy
n-d cross section, thought they play an important role at higher energies
and could explain some anomalies in polarization observables ($A_y$) \cite{GWHKG_96,WGHGK_PRL_98}.

The failure in the zero energy region can be corrected by including TNI.
In \cite{CCG_NPA_98,CCG_PLB_99} an hyperadial TNI was added 
with parameters adjusted to ensure B$_3$=8.48 and B$_4$=29.0 MeV, a value
which takes into account the Coulomb correction in $^4$He.
In \cite{VRK_PRL_98} the much more elaborate Urbana VIII and IX forces were
included leading to $B_3=8.48$, $B_4=28.3$ MeV.
The scattering length obtained in this way are displayed in lower part of Table~\ref{A_nt}. 
The small differences come essentially from the slighlty different $B_4$ values to which they were adjusted.
The values for the MT I-III model potential \cite{MT_NPA_69} are also given and found
to be very close to the realistic NN+NNN interactions.
In view of that it seems -- at least in what concerns bound and zero energy states --
that the only role of TNI is to ensure the physical values for $B_3$ and $B_4$. 
If they lead to very close results despite their severe analytical differences
one can hardly pretend to learn something about them in such kind of calculations alone.
In practice they provide enough parameters to fit one number.

Figure~\ref{nt_S} shows the n-$^3$H cross section calculated including only the $J^{\pi}=0^+$ and $1^+$ sates. 
Experimental values are taken from \cite{PBS_PRC_80,SKS_60}.
Results obtained with the NN forces alone are in dot-dashed curve. 
Those including TNI are in solid line (separate contributions in long- and short-dashed lines)
and provide an accurate cross section until $T_{lab}\approx 0.5$ MeV.


\vspace{-0.4cm}
\begin{table}[htb]
\caption{Experimental results on n-$^3$H scattering length.}\label{A_exp}
\newcommand{\m}{\hphantom{$-$}}
\newcommand{\cc}[1]{\multicolumn{1}{c}{#1}}
\renewcommand{\tabcolsep}{1.2pc} 
\renewcommand{\arraystretch}{0.8} 
\begin{tabular}{@{} cccl }\hline
$a_c$            &   $a_0$        &  $a_1$	    &  Ref.	 \\ \hline
$3.68\pm0.05  $  & $3.91\pm0.12 $ & $3.6\pm0.1    $ &  \cite{SBP_PLB_80}	   \\
$3.82\pm0.07  $  & $3.70\pm0.62 $ & $3.70\pm0.21  $ &  \cite{HRCK_ZPA_81} 	\\
$3.59\pm0.02  $  & $4.98\pm0.29 $ & $3.13\pm0.11  $ &  \cite{RTWW_PLB_85}   I  \\
$3.59\pm0.02  $  & $2.10\pm0.31 $ & $4.05\pm0.09  $ &  \cite{RTWW_PLB_85}   II \\
$3.607\pm0.017$  & $4.453\pm0.10$ & $3.325\pm0.016$ &  \cite{HDSBP_PRC_90}  \\
\end{tabular}
\end{table}
\vspace{-0.5cm}

\begin{figure}[hbtp]
\begin{center}\epsfxsize=14cm \epsfysize=6.5cm\mbox{\epsffile{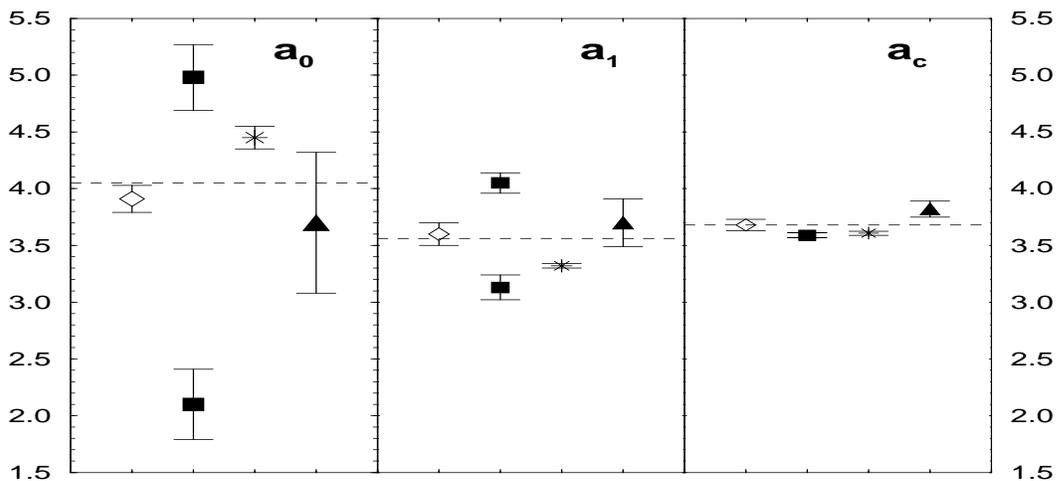}}\end{center}
\vspace{-1.0cm}
\caption{Diamond from \cite{SBP_PLB_80}, triangles up from \cite{HRCK_ZPA_81}, 
squares from \cite{RTWW_PLB_85}, stars from \cite{HDSBP_PRC_90}}\label{A_exp_fig}
\end{figure}

If the very low energy cross section is accurately measured and reproduced, the situation
with scattering lengths -- summarized in Table~\ref{A_exp} -- looks more precarious.
These values are displayed in Figure~\ref{A_exp_fig} together with the theoretical
ones previously discussed (horizontal lines).

\vspace{-1.0cm}
\vspace{-0.5cm}

The best agreement is found with the results of \cite{SBP_PLB_80}; 
in fact they contain a theoretical input, the ratio $a_1/a_0$, 
which turns to be very close to the one given by the realistic potentials from Table~\ref{A_nt}.
The other compatible results are those of \cite{HRCK_ZPA_81}. 
However, apart from the quite comfortable error bars in $a_0$, 
they have been obtained using a value of $a_c=3.82$ which is not
compatible with the more recent and precise values of \cite{RTWW_PLB_85,HDSBP_PRC_90}. 
The values given in \cite{HDSBP_PRC_90} are quite close to the theoretical ones
but they are extracted from a p-$^3$He R-matrix analysis in which the Coulomb interaction has been removed.
Finally, as it was pointed out in \cite{VRK_PRL_98},
the experimental values did not lie on the theoretical curves relying $a_i$ to $B_3$. 

The usual way  to get $a_i$ is by reversing the relations giving $\sigma(0)$ and $a_c$.
This procedure is numerically quite unstable.
Indeed by assuming an exact value $a_c=3.60$, the small existing error in $\sigma(0)$ leads to a range of values 
$a_0=4.60-5.16$ and $a_1=3.08-3.27$.
A more precise measurement of $\sigma(0)$ could be helpful to improve the present situation
and in this respect the CERN TOF \cite{CERN_TOF} neutron facility could offer interesting possibilities.

\bigskip
The preceding results show that the resonance peak requires the inclusion of
negative parity $J^{\pi}=0^-,1^-,2^-$ n-$^3$H states, which become dominante already at $T_{cm}\approx 2.5$ MeV.
 
\begin{figure}[htb]
\begin{minipage}[t]{79mm}
\epsfxsize=7.5cm\epsfysize=7.cm\mbox{\epsffile{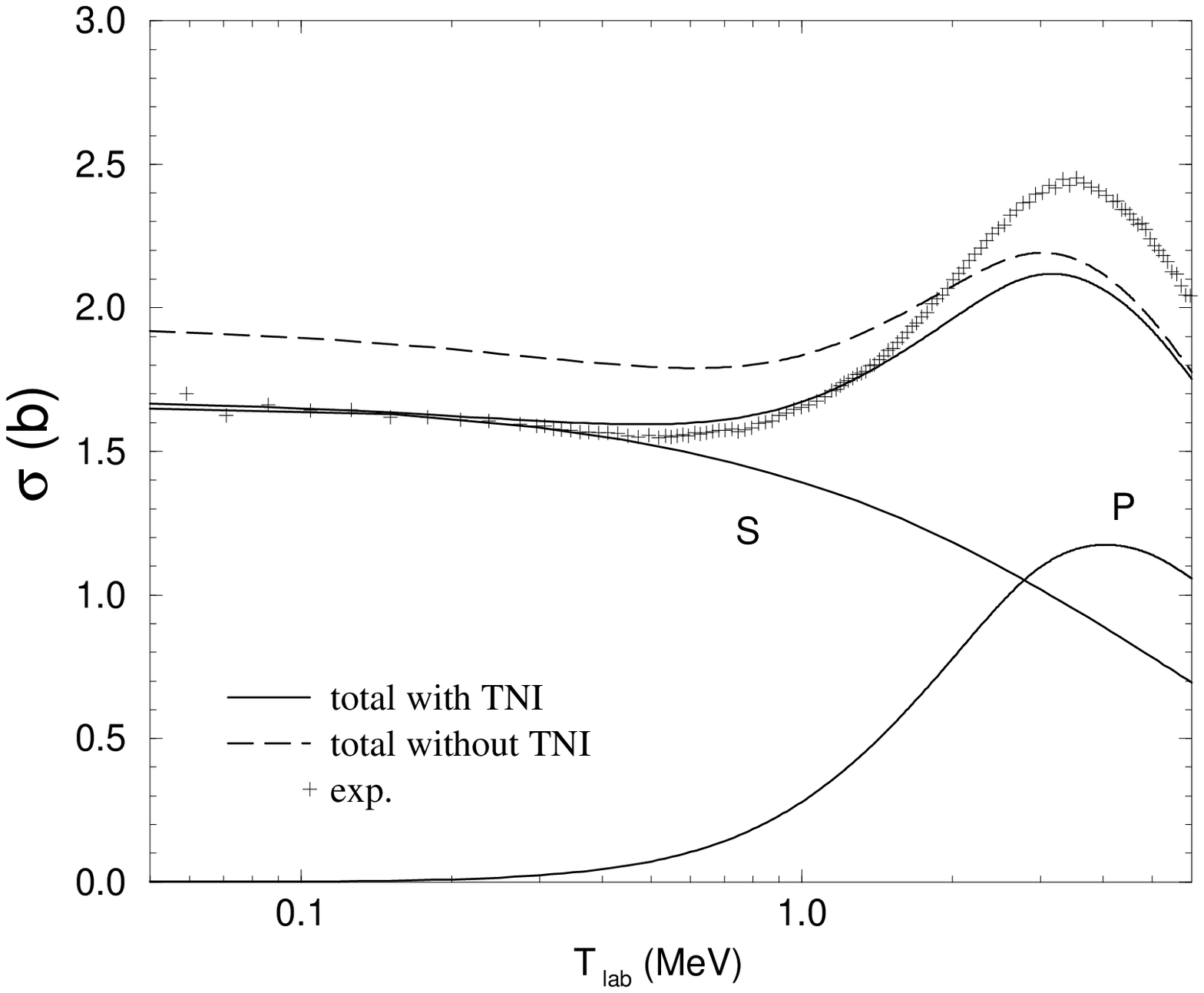}}
\vspace{-0.85cm}
\caption{S+P wave n-$^3$H cross section}\label{nt_tni_sp}
\end{minipage}
\hspace{0.2cm}
\begin{minipage}[t]{79mm}
\epsfxsize=7.5cm\epsfysize=7.cm\mbox{\epsffile{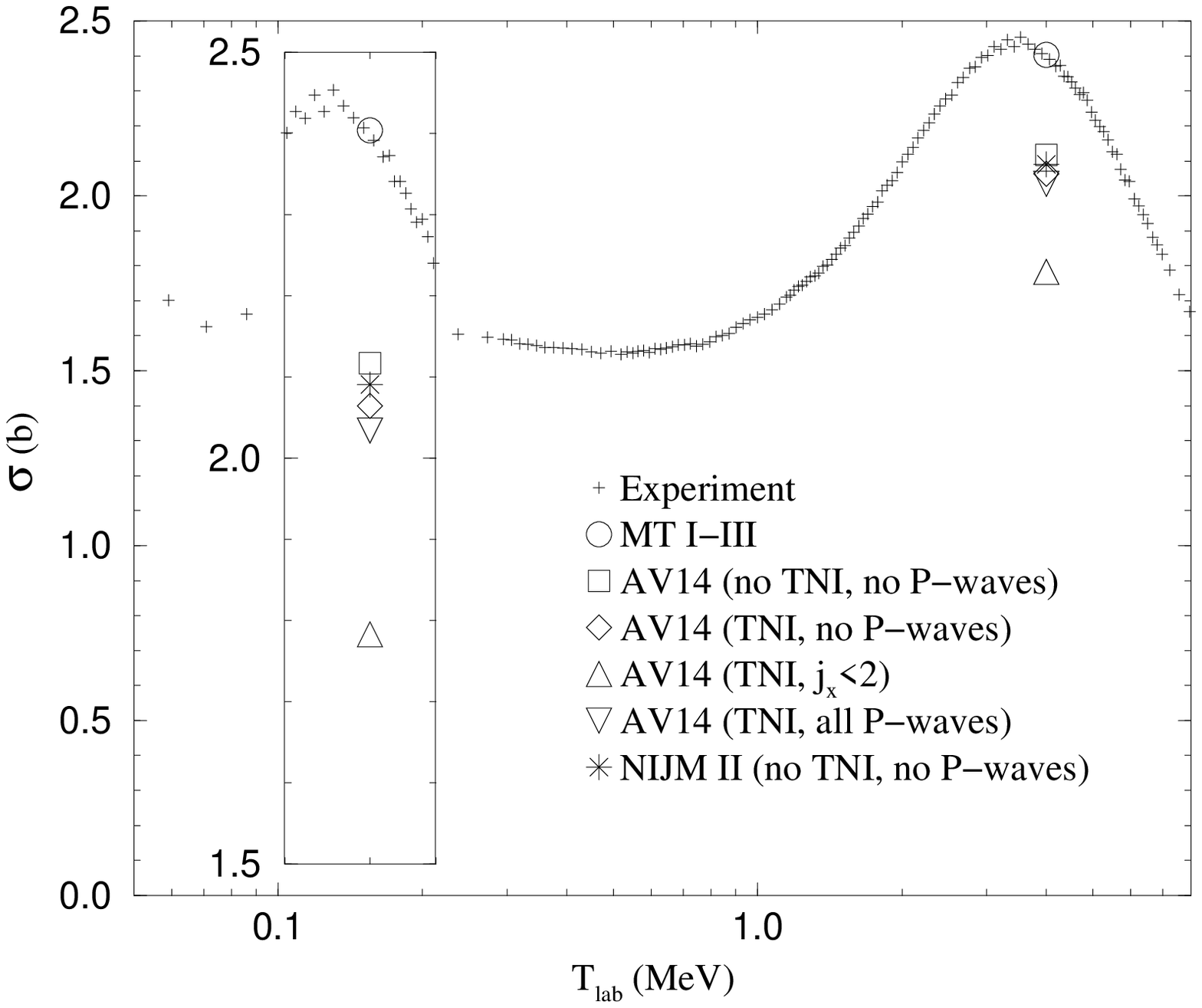}}
\vspace{-0.85cm}
\caption{NN P-waves at $T_{cm}$=3.5 MeV}\label{zoom_nt_tni_sp}
\end{minipage}
\end{figure}

A first attempt was done in \cite{CCG_NPA_98} using FY equations in configuration space.
The interation was limited to $V_{NN}^{j\le1}$+$^3$PF$_2$ 
and the partial wave expansion of FY amplitudes to $l_{y,z}\le2$.  
Results obtained with AV14 interaction are shown in Figure~\ref{nt_tni_sp}.
It was found that the peak region was poorly described by
the NN forces alone and that the inclusion of hyperadial TNI still lowered the cross section.
The calculations show a high sensitivity to the inclusion of P-waves in $V_{NN}$, 
a fact also pointed out by Fonseca in dd-dd and dd-p$^3$H polarization observables \cite{AF_NPA_98,AF_FBS_99}.
Their effect is shown in Figure~\ref{zoom_nt_tni_sp} with a zoom at $T_{cm}$=3.5 MeV:
including  $V_{^1P_1,^3P_0,^3P_1}$  still reduces
the cross section and the $V_{^3PF_2}$ rises substantially the value to compensate this reduction
but not enough to fit the experimental points.
This failure was attributed  either to a lack of convergence in the partial
wave expansion of  FY amplitudes or to a failure in NN current models \cite{CCG_PLB_99}.

\vspace{-0.5cm}
\vspace{-0.5cm}

A more recent calculation was done by Fonseca using AGS equations in momentum space \cite{F_PRL_99}.
This calculation is restricted to 1-rank separable expansion in the T$^{NN}$-matrix but
the partial wave expansion of AGS amplitudes was pushed until $l_{y,z}\le3$,
what represent a sizeable increase in the number of FY amplitudes,
specially those describing the internal structure of triton. 
Using AV14 and Bonn-B $V_{NN}$ models,
this author found a reasonable description of the total and differential cross section data.
The $V_{NN}$ P-waves -- essentially $^3$PF$_2$ -- make all the difference. 

This interesting result deserves some comments.
First to remark the unusually high sensitivity to NN P-waves in the low energy cross section.
Whereas it has almost no effect in the Nd scattering and triton binding energy,
the only contribution of the $V_{NN}^{^3P_2}$ accounts at $T_{cm}$=3.5 MeV 
for half of the n-$^3$H P-waves cross section,
which  in its turn  represents half of the total cross section .
Second to notice that the analyzing power $A_y$ shows
the same kind of discrepancy than for n-d but not solved 
by small changes in the $^3P_j$ NN phaseshifts, as done in \cite{TWK_PRC_98}.
Similar disagreements are also found in the dd-p$^3$H reaction.

It is worth noticing that -- despite its a priori crude approximation -- 
the 1-rank T-matrix expansion provides very close results
to those obtained by solving FY equations for the same number of amplitudes \cite{CCGF_FBS_99}. 
If the result  concerning the n-$^3$H resonant cross section is confirmed by independent methods
or by increasing the number of terms in the T-matrix,
it would speak very much in favour of low rank separable expansions.
 
\vspace{-0.80cm}
\begin{figure}[hbtp]
\begin{center}
\epsfxsize=12cm\epsfysize=8cm\mbox{\epsffile{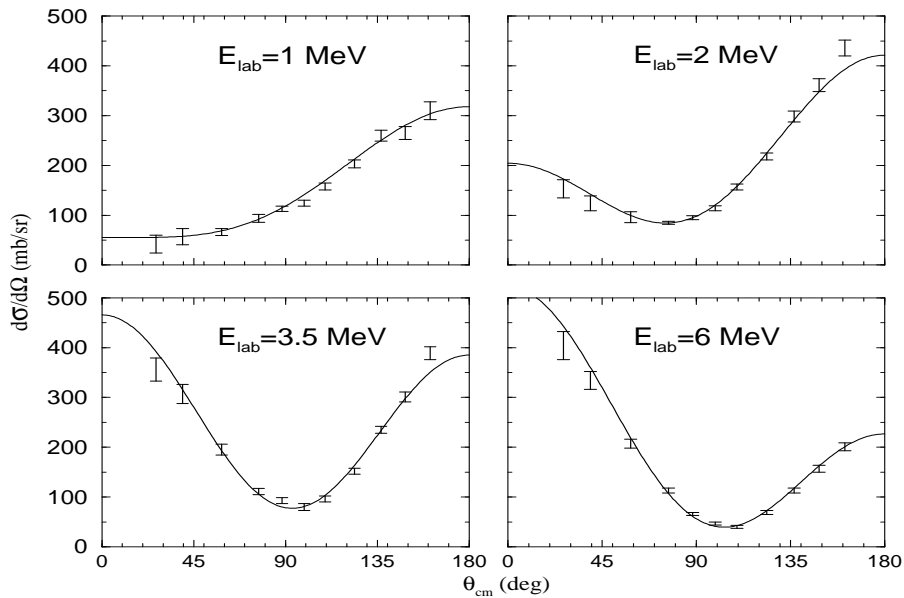}}
\end{center}
\vspace{-1.2cm}
\caption{n-$^3$H differential cross section with MT I-III  model}\label{dcs_nt_mt13}
\end{figure}
\vspace{-0.80cm}

It could have some interest to notice the ability of  
a trivial NN model like MT I-III in describing such a non trivial thing.
This potential acts only in $V_{NN}^{L=0}$ waves, has
no tensor, nor spin-orbit forces, even not pion tail and
triton wavefunction contains only S-wave Faddeev components.
It provides however a very good agreement with experimental results, 
specially in the resonance peak \cite{CC_PRC_98} and even for differential
cross sections, as can be seen in Figure~\ref{dcs_nt_mt13}.
Only the zero energy is slightly overestimated due to 
small differences in binding energies: $B_3=8.53$ MeV instead of $8.48$ and 
$B_4=30.3$ MeV instead of $29.0$, once removed Coulomb corrections.
In this model the n-$^3$H resonant cross section has nothing to do with NN P-waves:
it is created by the exchange mechanism between the incoming and target nucleons, 
what results into an effective 1+3 potential generated only by S-wave NN interactions.
This shows that nothing is trivial beyond A=2 and
the difficulty to disentangle the NN from the N-A interaction. 

\vspace{-.0cm}
\begin{table}[htb]
\caption{Experimental and theoretical values for p-$^3$He scattering length.}\label{a_p3He}
\newcommand{\m}{\hphantom{$-$}}
\newcommand{\cc}[1]{\multicolumn{1}{c}{#1}}
\renewcommand{\tabcolsep}{0.8pc} 
\renewcommand{\arraystretch}{0.90} 
\begin{tabular}{llllll}\hline
NN        & NNN        & Method & $a_0$        &  $a_1$        &	  Ref.                \\ \hline		 
MT I-III  &      ---   & CHH    &  10.0	     & 		   &    \cite{VKR_FB94}	    \\			 
          &            & CR     &  8.2	     & 7.7		   &    \cite{YF_FBS_99}    \\			 
AV18      &      ---   & CHH    & 12.9	     & 10.0  	   &    \cite{VRK_PRL_98}   \\			 
AV14      &Urbana VII  & VMC    &		     & 10.1$\pm$0.5  &    \cite{CRSW_PRC_91}  \\			 
AV14      &Urbana VIII & CHH    &  10.3	     & 9.13  	   &    \cite{VKR_FBS_95}   \\			 
AV18      &Urbana IX   & CHH    &  11.5	     & 		   &	  \cite{VRK_PRL_98}   \\ \hline			 
          &            & Exp.   & 10.8$\pm$2.6 &  8.1$\pm$0.5  &    \cite{AK_PRC_93}    \\			 
          &            &        &		     & 10.2$\pm$1.5  &    \cite{TB_AJ_83}	    \\			 
\end{tabular}
\vspace{-1.cm}
\end{table}

From the strong interaction point of view, the natural partner of n-$^3$H is p-$^3$He 
which differs only by Coulomb force.
The p-$^3$He reactions are however much more accessible experimentally 
though the resonant behaviour is somehow hidden due to the absence of total cross section.
The situation concerning the low energy parameters is summarized in Table~\ref{a_p3He}.
One can remark a much bigger theoretical "dispersion"  than for the n-$^3$H case
and a need for precise experimental values of $a_0$ and $a_1$.
Triplet scattering length is a relevant quantity in calculating the weak proton capture
p+$^3$He$\rightarrow^4$He+e$^+$+$\bar\nu$ cross section.
A firmly established $a_1$ value is needed 
but here -- as in many other radiative processes -- the main uncertainties come from the transition operators.
The MEC contribution can modify the result by a factor 5 \cite{CRSW_PRC_91}.
There exists also some preliminary calculations at non zero energy \cite{V_FBS_99}.
The differential cross sections at $E_{cm}=3$ MeV shows some lack at backward scattering angles 
and a rather large underprediction in the analyzing power $A_y$.

\section{The $^4$He continuum}

\begin{figure}[htb]
\begin{minipage}[t]{69mm}
\epsfxsize=6.cm\epsfysize=6.2cm\mbox{\epsffile{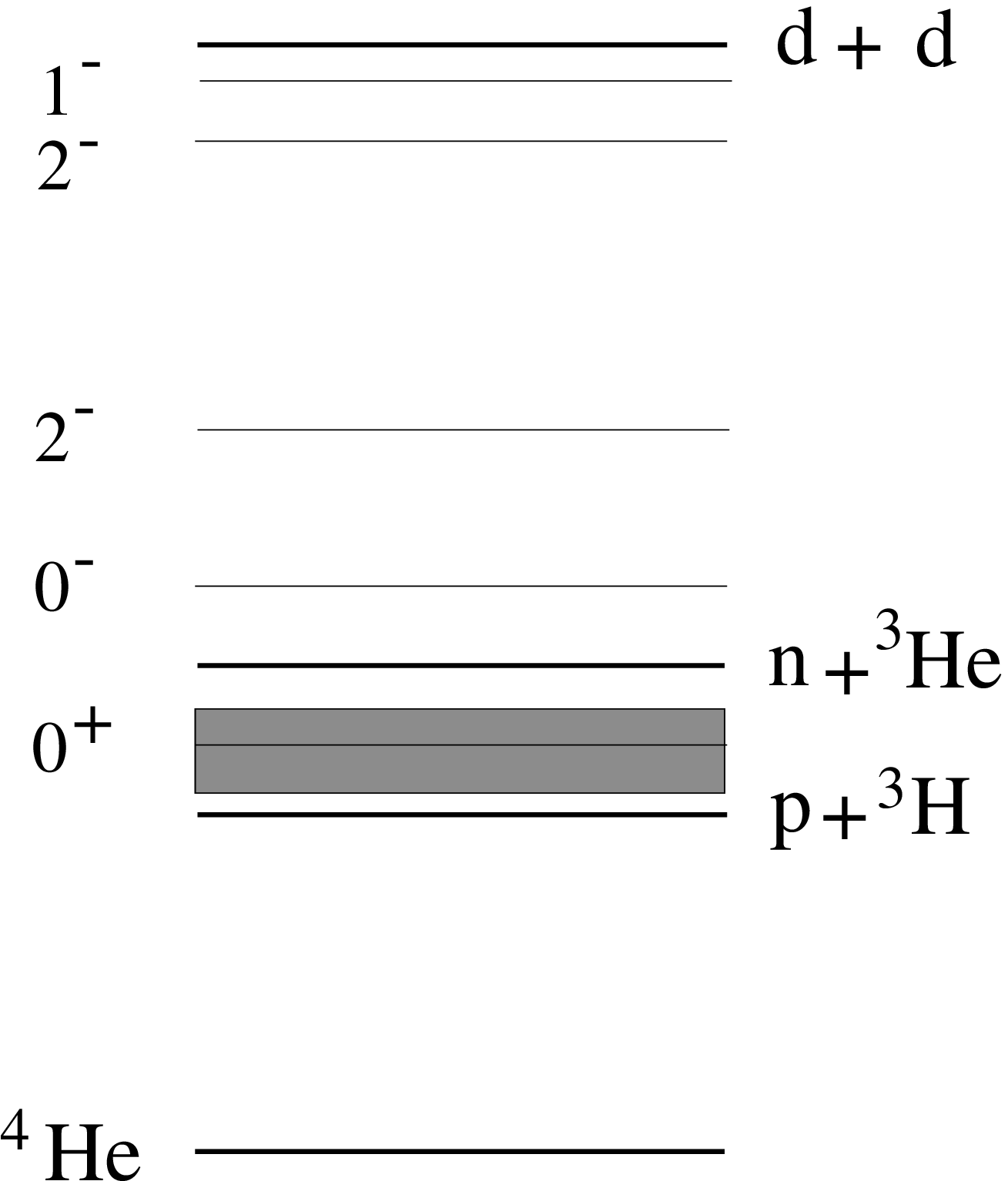}}
\vspace{-0.6cm}
\caption{$^4$He continuum spectrum}\label{He_Chart}
\end{minipage}
\hspace{0.5cm}
\begin{minipage}[t]{89mm}
\epsfxsize=8.8cm\epsfysize=6.5cm\mbox{\epsffile{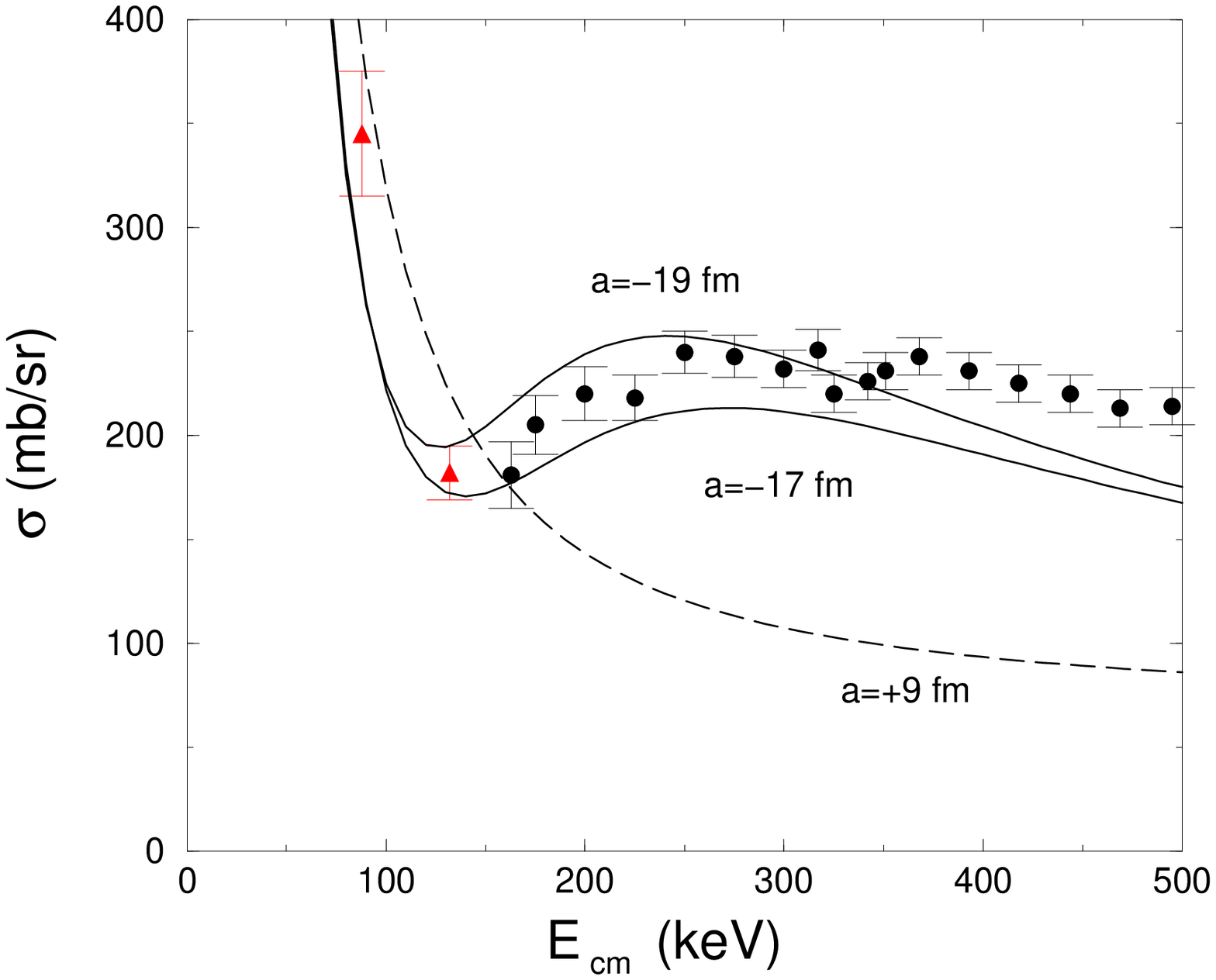}}
\vspace{-0.90cm}
\caption{p-$^3$H differential cross section at $\theta_{cm}=120^{\circ}$ 
versus center of mass energy}\label{pt_MT13}
\end{minipage}
\vspace{-.5cm}
\end{figure}

Next in complexity is the continuum spectrum of the $^4$He represented in Figure~\ref{He_Chart}.
Calculations of p-$^3$H scattering are complicated by the existence
of the first $J^{\pi}=0^+$ excitation of $^4$He in its threshold vicinity. 
This resonance located at $E_R$=0.40 MeV above  p-$^3$H 
covers with its $\Gamma$=0.5 MeV width the scattering region below n-$^3$He.
\vspace{-1.cm}

It turns out that most of the calculations performed until now find this state below the p-$^3$H
threshold, that is as a second $^4$He bound state, probably
because  they did not include Coulomb corrections.
Due to that, the sign of the strong p-$^3$H scattering length is wrong and the interference with the Coulomb
amplitude leads to senseless results in the interthreshold region.

\begin{table}[htb]
\vspace{-.5cm}
\caption[]{Low energy N+3N parameters (fm)}\label{tab_lep}
\newcommand{\m}{\hphantom{$-$}}
\newcommand{\cc}[1]{\multicolumn{1}{c}{#1}}
\renewcommand{\tabcolsep}{1.2pc} 
\renewcommand{\arraystretch}{1.0} 
\begin{tabular}{ccrrrr}
S&T&   $a$    &  $r_0$      & $v_0$          & $q_0$  \\ \hline
0&0& 14.75    &  6.75       & 0.308          & 0.505	            	 \\
0&1&  4.13    &  2.01       & 0.462          & -	            	  \\
1&0&  3.25    &  1.82       & $\simeq$ 0     & -	            \\
1&1&  3.73    &  1.87       & 0.231          & -                  	  
\end{tabular}
\vspace{-.6cm}
\end{table}

Among the few data existing in this energy range, we found the differential cross section
at $\theta_{cm}=120^{\circ}$ as a function of the energy in \cite{JSSL_PR_63}.
An attempt to describe this cross section can be done using the low energy scattering parameters
given in \cite{CC_PRC_98} and summarized in Table~\ref{tab_lep}.
One infers from them, in the isospin approximation, the strong p-$^3$H scattering lengths $a^{S=0}=9.44$ and $a^{S=1}=3.49$ fm.
The  p-$^3$H  amplitude reads $f(\theta)=f_c(\theta)+f_{sc}(\theta)$ where $f_c$ is the pure Coulomb
term and $f_{sc}$ the strong amplitude in a Coulomb field \cite{GW}. 
Limiting $f_{sc}$ to S-wave and using the Coulomb corrected effective range approximation, 
one can estimate the low energy p-$^3$H differential cross section.
The results obtained with the values of Table~\ref{tab_lep} and an effective range $r_0=2$ fm
for both spin states are displayed in Figure~\ref{pt_MT13} (dashed line) and
fail to reproduce the observed structure.
A good fit (solid lines) is obtained with large and negative values $a^{S=0}\approx-20$ fm.
indicating the  above threshold position of the first $^4$He excitation.
In terms of isospin components, this corresponds to a value $a^{S=0}_{T=0}\approx-40$.
The precise location of this state is a very strong requirement for the NN models.
Without it, there is no hope to have a good description of the low energy p-$^3$H and also n-$^3$He reaction. 
We remark that a direct CRM calculation  was presented in \cite{YF_FBS_99}.  
These authors found $a^{S=0}=-22.6$ fm, $a^{S=1}=4.6$ fm and extracted from the 
p-$^3$H phaseshifts the position $E_R=0.15$ MeV and width $\Gamma=0.23$ MeV of the resonance. 

There exists some calculation for the n-$^3$He scattering length. 
Using MT I-III and CRM, \cite{YF_FBS_99} found $a_0=7.5+4.2i$ and $a_1=3.0+0.0i$.
Using AV14 and Urbana VII TNI, \cite{CRSW_PRC_90} found $a_1=3.5\pm0.25$, in agreement with the experimental
value although the coupling to p-$^3$H was neglected. The zero energy wavefunction
was used to obtain the n+$^3$He$\rightarrow^4$He+$\gamma$ cross section
which was found overpredicted by a factor 2 and 1.5 in \cite{SWPC_PRC_92} where the $\Delta$
degree of freedom was included.
Results for the n-$^3$He elastic cross section at higher energy have been presented in this conference \cite{Uzu_FB16}.
None of these calculations have been performed taking into account the full complexity required.

Several d-d calculations have been performed at different levels of approximation.
Using VMC authors of \cite{APS_PRC_91} calculate the d+d$\rightarrow^4$He+$\gamma$ 
cross section with a dd wavefunction decoupled
from n-$^3$He and p-$^3$H channels and conclude that the optimal variational 
wavefunction does not explain the data.
FY equations in momentum space and separable potential  were used in \cite{UOT_93,UOT_FBS_95,UKOT_97}   
to calculate total and differential $\vec{d}+\vec{d}\rightarrow$p+$^3$H 
cross sections at 20-120 keV and found very good agreement with data.
Also at threshold energies, FY equation in configuration space have been
solved using simple MT I-III  model and isospin approximation
for the N+NNN thresholds. A strong J=$0^+$ T=0  dd scattering length $a_0=4.91-0.011i$ fm was found. 
Using CRM with the same interaction and Coulomb taken into account 
authors of \cite{YF_FBS_99} found $a_0=10.2-0.2i$ and $a_2=7.5$ fm.
At higher dd energies, there exists AGS calculations  
of polarization observables indicating large disagreement with data \cite{F_PRL_99}.
Finally we would like to mention the extensive RGM calculations
of $^4$He bound and scattering states done by \cite{HH_NPA_97} in which the phase shifts
of different two-fragment channels were obtained.

\section{SUMMARY}

The continuum of A=4 is an open door to a higher degree in the nuclear complexity
and offers an enormous field of work for a generation of motivated researchers.

The n-$^3$H resonant peak acts as a zoom for the internal structure of triton
as well as for the NN P-waves. A first task is to clarify with independent calculations
the ability of NN models in describing the elastic total and differential cross section
in the resonance region. The position and width of the underlying resonances
could be then calculated. 
The results obtained using the complex rotation method in  A=3 system seems very promising \cite{KMH_FB16_00}.

A major point is the description of the p-$^3$H and n-$^3$He thresholds,
dominated by the first excitation of $^4$He. 
This calculations imply coupled channel four-body equations with Coulomb interaction taken into account.
This structure completed with the d-d channel constitutes the formal skeleton
to be acquainted with in order to access the A=4 continuum.

The calculation of the numerous weak and electromagnetic capture processes is a redoubtable challenge.
The access to a high quality of nuclear wavefunction
would allow to test and fix the many ambiguities in the transition operators.

The last remark concerns relativity. The increasing complexity of the three nucleon forces 
requires a numerical investment which becomes comparable to a relativistic description.
An effort, conceptual and numerical, in this direction should be of highest interest.
The pioneering result using Gross equation \cite{GS} shows the possibility
to reach a consistent description of the three nucleons system using only NN forces.
The description of a nuclear system will probably remain
always phenomenological -- nucleons are already very complicated objects --
but the relativistic approach could provide a simpler framework.

\bigskip
{\bf Acknowledgments}:
The author is deeply grateful to A. Fonseca, H. Hofmman, A. Kievsky, S. Oryu, E. Uzu, B. Pfitzinger and M. Viviani for helpful 
discussions and remarks during the preparation of this contribution
and to M. Mangin-Brinet for a carefull reading of the manuscript.

\end{document}